\documentclass[10pt,letterpaper]{article}
\usepackage{opex3}
\usepackage{color}

\usepackage{amsmath,amssymb,graphicx}
\usepackage{epsfig}
\usepackage{epstopdf}
\usepackage{verbatim}

\newcommand{\abs}[1]{\left| #1 \right|} 
\newcommand{\be}{\begin{equation}}
\newcommand{\ee}{\end{equation}}
\newcommand{\myeq}[1]{Eq.~(\ref{eq:#1})}

\begin{document}

\title{Paraxial full-field cloaking}

\author{Joseph S. Choi$^{1,*}$ and John C. Howell$^{2}$}

\address{$^1$The Institute of Optics, University of Rochester, Rochester, New York 14627, USA\\
$^2$Department of Physics and Astronomy, University of Rochester, Rochester, New York 14627, USA}

\email{$^*$joseph.choi@rochester.edu} 

\begin{abstract}
We complete the `paraxial' (small-angle) ray optics cloaking formalism presented previously [Choi and Howell, \opex \textbf{22}, 29465 (2014)], 
by extending it to the full-field of light.  Omnidirectionality is then the only relaxed parameter of what may be considered an ideal, broadband, field cloak.
We show that an isotropic plate of uniform thickness, with appropriately designed refractive index and dispersion, can match the phase over the whole visible spectrum. 
Our results support the fundamental limits on cloaking for broadband vs. omnidirectionality, and provide insights into when anisotropy may be required.
\end{abstract}



\section{Introduction}
Invisibility efforts by scientists have focused on building a broadband, omnidirectional cloaking device~\cite{Gbur-Cloak-2013}.  Such a cloak that works in macroscopic dimensions, for the entire visible spectrum, with the full-field of light may be considered an `ideal' cloaking device~\cite{Fleury-Alu-2014}.  Much of this research was spawned by the initial omnidirectional cloaking designs, which used artificial materials (called `metamaterials') to control electromagnetic fields/waves~\cite{Pendry-2006,Leonhardt-2006}.  This field was called `transformation optics,' since coordinate transformations were used to design the necessary material properties that morphed space for light.  Transformation optics has produced many ingenious designs and devices~\cite{McCall-2013}.  However, creating the required spatial distribution of metamaterials, their anisotropy, and causality, limit the spectrum to single frequencies, or to a narrow bandwidth, and is particularly difficult to manufacture for visible frequencies~\cite{Gbur-Cloak-2013,McCall-2013}. 

Much of the scientific work on cloaking has retained the omnidirectionality and full-field (amplitude and phase of light waves) nature of transformation optics~\cite{Fleury-Alu-2014}.  
With this, to make forays into large bandwidths in the visible spectrum, researchers have built reflecting `carpet cloaks'~\cite{Li-Pendry-2008,Ergin-2010}, proposed non-Euclidean geometry mapping~\cite{Leonhardt-2009}, demonstrated macroscopic ray optics cloaks (matching directions, rather than phases, of fields)~\cite{Howell-Cloak-2014,Chen-2013}, and suggested strongly diamagnetic superconductors~\cite{Monticone-Alu-2014}. 

Simultaneously, researchers have investigated the fundamental limits on bandwidths for these omnidirectional cloaks.  It is generally agreed that causality requires such cloaking of a non-zero volume, either passive or active~\cite{Ma-2013}, to be only possible for single frequencies~\cite{Miller-2006}. 
Chen et al. showed that zero scattering cross-section of a cloaked object (for an omnidirectional transformation optics cloak in two-dimensions (2D)) cannot be attained by neighboring frequencies of the target frequency, no matter how narrow the bandwidth~\cite{Chen-PRB-2007}.  They then derived an upper bound for the bandwidth that was proportional to the scattering cross-section radius.
Hashemi et al. have also provided a theoretical bound for transformation optics cloaks, even if perfectly manufactured, which shows increased imperfections for increasing bandwidth~\cite{Hashemi-2012}.  
Derived from causality, their ``diameter-bandwidth product'' limit states that the effective bandwidth  is inversely proportional to the cloaked object diameter.  
They suspect a similar sensitivity to `carpet cloaks' as well.
Similarly, Monticone and Al\`{u} showed that any linear, passive, non-diamagnetic cloak must increase the overall scattering, compared to an uncloaked object, when integrated over all wavelengths~\cite{Monticone-Alu-2013}.  Even if scattering is suppressed in a finite bandwidth, this will need to be ``paid back'' in the rest of the spectrum.
On the other hand, Leonhardt and Tyc's broadband solution is a three-dimensional (3D), anisotropic cloak~\cite{Leonhardt-2009}.  Their ``price to pay for practical invisibility'' is relaxing the full-field requirement to allow time delays.

Our approach has been to begin with broadband cloaking for the entire visible spectrum, but within the paraxial approximation, then to expand the cloaked size and angles~\cite{Discovery-CA-2015}.  This is typically the method used in optical engineering, as increasing the field-of-view (angles) and/or numerical aperture (refractive index multiplied by the solid angle of the accepted light) can be challenging~\cite{Smith-LensDesign-2005}.  
So in this paper, we use the definition for ``cloak'' that is to ``hide,'' since the cloaked object is not necessarily covered in all directions as if by a garment.
Our previous ray optics demonstration relaxed both angles and phase to show that only four lenses were needed for 3D, macroscopically scalable, stand-alone cloaking for small angles~\cite{Choi-Cloak1-2014}.  Our paraxial cloak was the simplest design we found that used isotropic optics designed for the entire visible spectrum.  These optics have been well understood for centuries~\cite{Born-Wolf-2010} and are readily available off-the-shelf.
Interestingly, Hashemi et al. concluded by suggesting a relaxing of angles and phase, to possibly obtain a precise understanding of when the cloaking problem becomes easy~\cite{Hashemi-2012}.  
They considered knowing what cloak was the ``weakest'' relaxed, yet still practical for large objects, to be valuable, due to the difficulties of transformation optics cloaks.

\section{Extending paraxial cloaking to include the full-field}
We now remove the ray optics requirement, to complete a paraxial cloaking theory that works for the full-field (matched amplitude \emph{and} phase).  Such paraxial full-field cloaking satisfies all but the omnidirectionality condition of an `ideal' cloaking device.
This `paraxial' formalism can work for up to $\pm 30^\circ$~\cite{Siegman-book-1986}, which is practical for many cases where the cloak is not placed immediately before an observer.
Although we discuss the propagation of a monochromatic field, 
since an arbitrary field of light can be written as a linear superposition of monochromatic waves, our theory extends to broadband without loss of generality.

Duan et al. recently provided phase matching with their unidirectional cloaking system based off of geometric optics~\cite{Duan-2014}.  They provided a heuristic reason why rays that pass through the edge of their optical system may match the phase of rays that pass through the center instead, for their particular setup with split lenses.  
They  then simulated phase-matched cloaking for discretely separated wavelengths of micro-waves to mm-waves.  
Here we analytically show phase matching for general paraxial optical systems, including continuously multidirectional cloaks.
It is based on the formula given by Siegman and others, for propagation of any paraxial field through a generalized paraxial optical system~\cite{Siegman-book-1986}.  The resulting formula is proved using Fermat's principle and Huygens' integral, by accounting for the optical path lengths of all rays.

Field propagation based off of Huygens' principle of wavelet propagation, is effective and widely used in diffraction theory and Fourier optics~\cite{Goodman-book-2005}.  We first assume that the ambient medium is spatially uniform with index of refraction $n$, and that the optical system is rotationally symmetric for simplicity. 
Non-uniform medium~\cite{Choi-Cloak1-2014} or non-rotationally symmetric systems~\cite{Siegman-book-1986} can be derived from here.
\begin{figure}
\begin{centering}
\includegraphics[scale=1]{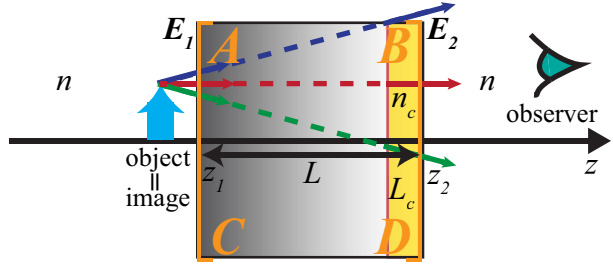}
\caption{\label{fig:CloakBox-ray2field}
\textbf{Ideal (`perfect') paraxial cloak.}
The image from a cloaking device is the same as the object.  
Propagation of light rays can be described by an `ABCD' matrix.  $n$ is the index of refraction of the ambient medium. 
$L$ is the longitudinal length of the device along the center $\mathbf{z}$-axis. 
$E_1, E_2$ are the input and output fields, respectively, at $z=z_1$ and $z_2$.
Phase matching is achieved with a flat plate with index $n_c$ and length $L_c$.
}
\end{centering}
\end{figure}

Additionally assuming no limiting apertures, Huygens' integral in the Fresnel (or, paraxial) approximation is given by~\cite{Siegman-book-1986,Collins-1970} (See Fig.~\ref{fig:CloakBox-ray2field}):
\begin{align}
\tilde{E}_2(x_2,y_2) \!\! = \!\!
\frac{i e^{-i k_0 L_0}}{B \lambda_0}
\!\!\!\!\iint^\infty_{-\infty}
\!\!\!\!\tilde{E}_1(x_1,y_1 \!)
\exp \! 
\left\{
\!\!\!
-i \frac{\pi}{B \lambda_0}
\!\!\left[ 
A \! \left(x_1^2 \!\! + \!\! y_1^2\right) 
\!\! - \! 2 \! \left(x_1 x_2 \!\! + \!\! y_1 y_2\right) 
\!\! + \! D \! \left(x_2^2 \!\! + \!\! y_2^2\right) 
\right]
\!\!
\right\}
\!\!
\mathrm{d}x_1
\mathrm{d}y_1
\label{eq:Huygens-E2}
\!.
\end{align}
$A, B, C, D$ are the ABCD matrix coefficients.
$L_0 = \sum_i n_i L_i$ is the on-axis optical path length, where each $i^\text{th}$ optical element has index of refraction $n_i$ and physical thickness $L_i$ along the longitudinal axis ($\mathbf{z}$).
$\lambda_0$ and $k_0$ are the free space wavelength and wave vector, respectively.
$\tilde{E}_1, \tilde{E}_2$ are the complex, spatial amplitudes of the input and output field distributions, respectively ($E_1, E_2$ in Fig.~\ref{fig:CloakBox-ray2field} are their real parts, but without the $e^{+ i\omega t}$ harmonic time dependence).

In our previous work, we stated that a `perfect' cloaking device (of length $L$) simply replicates the ambient medium throughout its volume, 
and that its ABCD matrix is given by~\cite{Choi-Cloak1-2014} 
\begin{equation}
\begin{bmatrix}
A & B \\
C & D
\end{bmatrix}_{\text{perfect cloak}}
=
\begin{bmatrix}
1 & L/n \\
0 & 1
\end{bmatrix}
.
\label{eq:cloak-metric}
\end{equation}
Thus, the propagated field for a `perfect' full-field cloak, of length $L$, is:
\begin{align}
\tilde{E}_2^{\text{cloak}}(x_2,y_2) \!\! = \!\! 
\frac{i n e^{-i k_0 n L}}{L \lambda_0}
\!\!\!\! \iint^\infty_{-\infty}
\!\!\!\!  \tilde{E}_1(x_1,y_1)
\exp \!
\left\{
\!\! -i \frac{n \pi}{L \lambda_0}
\!\! \left[ 
\left(x_1^2 \!\! + \!\! y_1^2\right) 
\!\! - \!\! 2 \left(x_1 x_2 \!\! + \!\! y_1 y_2\right) 
\!\! + \!\! \left(x_2^2 \!\! + \!\! y_2^2\right) 
\right]
\!\! \right\}
\! \mathrm{d}x_1
\mathrm{d}y_1
\label{eq:E2-perfect-field-cloak}
\! .
\end{align}

A ray optics cloak satisfies \myeq{cloak-metric}. 
By comparing \myeq{Huygens-E2} and \myeq{E2-perfect-field-cloak}, we see that a ray optics cloak can be a full-field cloak, if $e^{-i k_0 L_0} = e^{-i k_0 n L}$.
Specifically, this is the case when $k_0 L_0 = k_0 n L$ (absolute phase-matching), 
or when $L_0 \equiv n L \pmod{\lambda_0}$ (phase-matching to integer multiples of $2\pi$).
The significance of these conditions is that they allow for phase-matched, full-field cloaking for \emph{any} incoming fields within the paraxial approximation.  
The phase-matching condition only needs to be satisfied once for a given optical system, and then all ray directions and positions, or any field distribution ($\tilde{E}_1$) will exit as if traversed through ambient space.
Note that a ray optics cloak will usually be phase-matched with a full-field cloak, by integer multiples of $2\pi$, for multiple, but discretely separated, wavelengths automatically. 
However, we can do better and match it for a continuous, broad bandwidth with appropriate dispersion control.

One method for broadband phase matching is to add a thin, flat plate to a ray optics cloak, anywhere between the background object(s) and the observer.  Since it is ``thin'' and flat, the original ABCD matrix (\myeq{cloak-metric}) will be unchanged~\cite{Handbook-Opt-v1-2010,FG-GO-2004} and only the $e^{-i k_0 L_0}$ factor outside the integral of \myeq{Huygens-E2} will be affected. 
We consider the case shown in Fig.~\ref{fig:CloakBox-ray2field}, where the flat plate is placed immediately after a ray optics cloak.
Let $L' = \sum_{i=1}^{N} L_i$ be the total length of the \emph{original} ray optics cloak (so $L=L_c+L'$), where $N$ is the number of original optical elements.
So $L_0 \rightarrow n_c L_c + \sum_{i=1}^{N} n_i L_i$ in \myeq{Huygens-E2}, and 
$n L \rightarrow n (L_c + L')$ in \myeq{E2-perfect-field-cloak}.
Here, $n_i, L_i$ are the index of refraction and physical longitudinal length, respectively, for the $i^{\text{th}}$ optical element, and  $n_c, L_c$ are likewise for the correcting thin plate.
It can be shown that a plate is ``thin'' when $\abs{ {L_c}/{L'} } \ll 1$ and $\abs{ {L_c}/{L'} } \ll \abs{ {n_c}/{n} }$.
For our thin correcting plate, its phase-matched dispersion relation is
\begin{align}
n_c(\lambda_0, m, L_c) 
=
n(\lambda_0) 
+
\frac{1}{L_c}
\left\{
m \lambda_0 +
\sum_{i=1}^{N}{ \left[
n(\lambda_0) - n_i(\lambda_0)
\right] L_i }
\right\}
\label{eq:thinplate-phase-correct-2-m}
.
\end{align}
$m$ is the integer multiple of $2\pi$ for which the phase is matched, with discrete but infinite choices.

\begin{figure}
\begin{centering}
\includegraphics[scale=1]{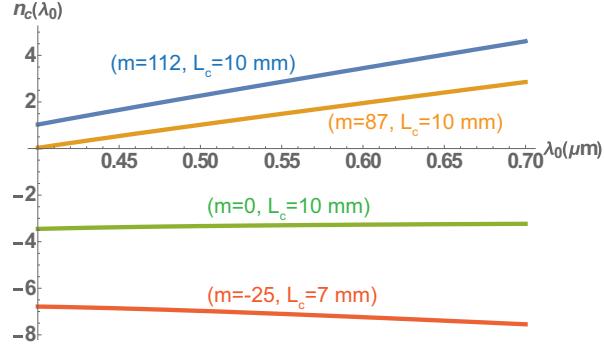}
\caption{\label{fig:4lens-nc-1}
\textbf{Dispersion of phase-correcting plate for a four lens symmetric cloak.}
Various thin, flat plates are used to match the full-field cloak condition (\myeq{E2-perfect-field-cloak}), based on the ray optics four lens cloak in \cite{Choi-Cloak1-2014}. 
The dispersion relation of the refractive indices for these plates are shown, for various values of $m$ and $L_c$, using \myeq{thinplate-phase-correct-2-m} in ambient air.
}
\end{centering}
\end{figure}

There are infinitely many combinations of $L_c$ and $m$ that are possible for $n_c(\lambda_0, m, L_c)$.  Ultimately, the choice of which to use should depend on physically realizable $n_c$'s.
A few such solutions, that phase-corrects the four lens ray optics cloak in \cite{Choi-Cloak1-2014}, are shown in Fig.~\ref{fig:4lens-nc-1}.
For simplicity, we assumed that the cloaking system is placed in air, with $n=n_\text{air}=1$ for all wavelengths, and we ignored any coatings.  The lens coatings can be modeled as additional thin plates as needed.
We also only consider the visible spectrum (400-700 nm, for our discussion).
A thicker compensating flat plate can reduce the required dispersion range, but this also affects the imaging quality, so we have purposely limited its thickness here.

Many of the solutions for positive refractive indices require anomalous dispersion, as shown in Fig.~\ref{fig:4lens-nc-1}.  For the purposes of cloaking, such dispersion for broadband spectrum with low losses is needed to imitate ambient space properly.
This is similar to the finite bandwidth cloak made of anisotropic layers, as suggested by Kildishev et al., which required strong anomalous dispersion combined with loss compensation~\cite{Kildishev-2008}.
Costa and Silveirinha have suggested using nanowire metamaterial to provide such anomalous dispersion, and they numerically calculated their index of refraction to be near $2.9 \sim 3.1$ for the entire visible spectrum~\cite{Costa-2012}.  
They can achieve these low loss, broadband, anomalous dispersion properties by utilizing the collective, spatial properties of metamaterials.  
This allows the high loss and narrow band properties of typical transparent materials, imposed by the Kramers-Kronig relations, to be overcome~\cite{Silveirinha-2009}.
In addition, Theisen and Brown have experimentally demonstrated anomalous dispersion for $0.5\sim 1$ $\mathrm{\mu m}$ wavelengths, with Gallium implanted Silicon pads~\cite{Theisen-FiO-2012}.  These have refractive indices near 2 or 4, with variations of about 1 over the spectrum, depending on the doping level.

Negative-index metamaterials are good candidates for phase-correction as well~\cite{Veselago-1968}.  The utilization of metamaterials may have been expected, since we expanded ray optics cloaking to the field cloaking aspect of transformation optics.  
Much progress is being made, both theoretically and experimentally, for creating negative index materials for broadband optical frequencies~\cite{Soukoulis-2011}.  
Some refractive index values demonstrated experimentally include 
between -3 and 0 for microwave frequencies in 2D~\cite{Shelby-2001}, 
n = 0.63  at 1,200  nm to n = -1.23  at 1,775  nm in a low loss, 3D bulk material~\cite{Valentine-NIM-2008}, and n = 1 to -7.5 for 1.1-2.4 $\mathrm{\mu m}$ wavelengths~\cite{Chanda-2011}.

\section{Discussion}
By relaxing only omnidirectionality for an `ideal' cloak, we have shown how to match the phase for the whole visible spectrum.  The phase-matching plate may require negative index metamaterials or anomalous dispersion, which are broadband and low loss, but current research has shown much progress in this regard.  
We had shown that building a 3D, broadband, macroscopic cloak, for the visible spectrum, can be fairly ``easy'' for ray optics in the small angle limit~\cite{Choi-Cloak1-2014}.  
Extending this to the full-field seems to not be too challenging with recently developed materials.  
As with typical lens designs, we expect that extending such cloaks to large angles may be difficult, though possible.  
However, making it work for all angles appears to likely be fundamentally limited.  
Realizing omnidirectionality from a paraxial full-field cloak has not been achieved to our knowledge, as this coincides with achieving broadband with transformation optics cloaks.  
By showing paraxial cloaking that is broadband to be practical, but without a similar ease for omnidirectional and broadband cloaks, our work supports recent work showing a trade-off between broadband and scattering cross-section.  
This is because small scattering cross-section implies large angles, and zero cross-section gives omnidirectionality.

It is interesting that paraxial full-field cloaking does not require anisotropy, though being 3D, macroscopic, and broadband.  
Anisotropy seems to be a requirement when creating an omnidirectional cloak, from a paraxial cloak, not necessarily a property of 3D or field cloaks alone~\cite{Pendry-2006,Leonhardt-2006,Leonhardt-2009,Greenleaf-2003}.
Although our cloak is broadband, Greenleaf et al. have shown that the cloaked material itself has eigenfrequencies that need to be avoided for proper cloaking~\cite{Greenleaf-CMP-2007,Greenleaf-NJP-2008}. 
As a side, most, if not all, cloaks to date, can be detected by measuring a pulse incident on the cloak~\cite{Fleury-Alu-2014,Leonhardt-2009}.  This is the same for our cloak presented here, unless absolute phase is matched with negative index materials.
Finally, an open question seems to be whether an isotropic, 3D, omnidirectional, broadband cloak can be achieved for ray optics.  
This is possible for anisotropic materials~\cite{Leonhardt-2009}, but an isotropic, 3D, omnidirectional, full-field cloak is not possible~\cite{Nachman-1988,Wolf-Cloak-1993}.
However, accepting time delays might allow some freedom, as shown here where ray optics allowed relaxing of the material requirements.

\section*{Acknowledgments}
The authors would like to thank Professor Allan Greenleaf for background on Calderon's inverse problem and uniqueness theorems, and Professor James Fienup for discussions on field propagation.
This work was supported by the Army Research Office Grant No. W911 NF-12-1-0263 and the DARPA DSO Grant No. W31P4Q-12-1-0015.
J.S.C was also supported by a Sproull Fellowship from the University of Rochester.

\end{document}